\documentclass{rmaa}

\usepackage{supertabular}

\title{Quantitative Stellar Spectral\\
       Classification. II. Early Type Stars}

\author{M. J. Stock\altaffilmark{1,2,3},
        J. Stock\altaffilmark{2},
        J. Garc\'{\i}a\altaffilmark{1}
    and N. S\'anchez\altaffilmark{1}}

\altaffiltext{1}{Laboratorio de Investigaciones Astron\'omicas,
                 Universidad del Zulia, Venezuela}
\altaffiltext{2}{Centro de Investigaciones de Astronom\'{\i}a
                 (CIDA), Venezuela}
\altaffiltext{3}{Postgrado de F\'{\i}sica Fundamental, Facultad de Ciencias
		Universidad de Los Andes, Venezuela}

\date{Received ...; accepted ...}

\fulladdresses{
\item M. Jeanette Stock, Javier Garc\'{\i}a, and N\'estor S\'anchez:
      Laboratorio de Investigaciones Astron\'omicas,
      Departamento de F\'{\i}sica, 
      Facultad Experimental de Ciencias,
      Universidad del Zulia,
      Apartado 15439, Maracaibo, Venezuela
      (mjstock@cida.ve,nestor@luz.ve,jgarcia@cata1.ivic.ve)
\item J\"urgen Stock:
      Centro de Investigaciones de Astronom\'{\i}a (CIDA),
      Apartado Postal 264, M\'erida 5101-A, Venezuela
      (stock@cida.ve)}

\shortauthor{M. J. Stock et~al.}
\shorttitle{Quantitative Stellar Spectral Classification. II.}

\resumen{
Se ha extendido, a las estrellas del tipo espectral B, el m\'etodo desarrollado
por Stock y Stock (1999) para estrellas del tipo espectral A-K, con el cual
es posible derivar magnitudes absolutas y colores intr\'{\i}nsecos a partir de los
anchos equivalentes de las l\'{\i}neas de absorci\'on de los espectros estelares.
Espectros de estrellas tipo-B  para las cuales el cat\'alogo de Hipparcos
proporciona paralajes con un error menor al 20\%, fueron observados con el
reflector de 1-m del CIDA equipado con espectr\'ografo Richardson y un detector
CCD Thompson 576x384. Utilizando una rejilla de 600 lineas/mm se obtuvo una
dispersi\'on en el primer orden de 1.753 \AA/pixel. Para cubrir el rango espectral
comprendido entre 3850 \AA\ y 5750 \AA\ fue necesario utilizar la rejilla en dos
posiciones distintas, teniendo un solapamiento en la regi\'on entre 4800 \AA\ y 4900 \AA. Fueron
observadas un total de 116 estrellas, pero no todas en las dos posiciones de
la rejilla.
Se identificaron un total de 12 lineas de absorci\'on en los espectros y se
midieron sus anchos equivalentes. Estos fueron relacionados con las magnitudes
absolutas derivadas del Cat\'alogo Hipparcos y con los colores intr\'{\i}nsecos (deducidos
de los tipos espectrales MK), por medio de polinomios de primer y segundo orden y
combinaciones de dos o tres lineas como variables independientes. Las mejores
soluciones fueron obtenidas con polinomios de tres lineas, reproduciendo
las magnitudes absolutas con un residuo promedio de 0.40 magnitudes, y los
colores intr\'{\i}nsecos con un residuo promedio de 0.016 magnitudes.}

\abstract{

The method developed by Stock \& Stock (1999) for stars of spectral
types A to K to derive absolute magnitudes and intrinsic colors from
the equivalent widths of absorption lines in stellar spectra is
extended to B-type stars. Spectra of this type of stars for which
the Hipparcos Catalogue gives parallaxes with an error of less than
20\% were observed with the CIDA 1-meter reflector equipped with a
Richardson spectrograph with a Thompson 576x384 CCD detector. The
dispersion is 1.753 \AA/pixel using a 600 lines/mm grating in the
first order. In order to cover the spectral range 3850 \AA\ to 5750 \AA\
the grating had to be used in two different positions, with an
overlap in the region from 4800 \AA\ to 4900 \AA. A total of 116 stars was oberved, but
not all with both grating positions.
A total of 12 measureable absorption lines was identified in the spectra and
their equivalent widths were measured. These were related to the
absolute magnitudes derived from the Hipparcos Catalogue and to the
intrinsic colors (deduced from the MK spectral types) using linear
and second order polynomials and two or three lines as independent
variables. The best solutions were obtained with polynomials
of three lines, reproducing the absolute magnitudes with an average
residual of about 0.40 magnitudes and the intrinsic colors with an
average residual of 0.016 magnitudes.}


\begin{document}
 
\maketitle                                                                      

\section{Introduction}

In a previous work two of the authors (Stock \& Stock,1999)
published a method for the derivation of stellar physical parameters
such as the absolute magnitude, an intrinsic color, and a metallicity
index from the equivalent widths or pseudo-equivalent widths of
aborption lines in stellar spectra. Use was made of a library of
stellar spectra made available by L. Jones (1999). Of these stars
those were rejected for which the Hipparcos Parallax Catalogue
gives parallax errors larger than 20\% of the parallax itself. Due
to this restriction the number of remaining O- and B-type spectra
was too small to be included. To close this gap we decided to launch
our own observing program dedicated exclusively to early type stars.
A number of recent papers have been dedicated to the subject
of a quantitative stellar classification. 
Malyuto \& Schmidt-Kaler (1999) classify G-, K-,
and M-type stars on the basis of spectral indices derived in the
region from 6000 to 10000 \AA. A similar aproach is used by 
Malyuto, Oestreicher and Schmidt-Kaler (1997) for K- and M- type stars based
on spectra in the region from 4800  to 7700 \AA.

\section{Observations}

In view of the restriction just mentioned we started an observing
program with the Richardson spectrograph of the CIDA observatory
attached to the 1-meter reflector, concentrated on O- and B-type
stars which fulfilled the same parallax error restriction used in
the previous work. A grating of 600 lines/mm was used in the first
order yielding a dispersion of 1.753 \AA/pixel. The detector
is a Thompson 576x384 CCD with a pixelsize of 23 microns. The
spectral range captured by a single exposure is about 1000 \AA.
Two grating positions were used, one yielding usable spectra from 3950 \AA\
to 4900 \AA, the other from 4800 \AA\ to 5750 \AA. In the following we
distinguish these as the B- and V-spectra. The grating positioning
mechanism was rather crude (it has now been improved) such that the
different spectra are not always centered at exactly the same
wavelength. For this reasom some of the lines near the end of the
spectra are not always covered.

The observing list was made up with the Bright Star Catalogue
in combination with the Hipparcos Parallax Catalogue, selecting all
O- and B-stars brighter than the 6th apparent V-magnitude. Of these
116 stars were observed, resulting in 72 stars with B-spectra only,
39 stars with B- and V-spectra, and 5 stars with V-spectra only.A
typical example of a B- and a V-spectrum is shown in Figure~1. Table~1
contains a list of all observed stars with their HD-numbers and 
additional pertinent information. An intrinsic color-magnitude
diagram of the observed stars is shown in Figure~2.

\section{Data reduction}

As in the previous paper the data analysis will be based on the
equivalent widths of absorption lines. When the true continuum is
resolved, and this is here the case, the determination of the
equivalent width is a standard procedure and need not be described
here. The signal/noise ratio naturally depends on the brightness of
the star, the exposure time, and a series of other factors.  In most
cases the value of S/N is found near 50.
A total of 11 measurable absorption lines was found in the
B-spectra, and only 4 in the V-spectra. Three of these are common
to both the B- and the V-spectra. Thus only one additional line was
added by including the V-band in the observing program. Even so the
inclusion of the V-band turned out to be quite fortunate as will be
seen in the analysis. In the first place it gave us a convenient
handle to determine the accuracy with which the equivalent widths
are determined. Also it turned out that the line added by the V-band
provides an important classification criterion.

For the lines in common to both bands we can carry out an analysis
of the accuracy with which the equivalent widths are determined. Only
three lines are available for the test. There is a pronounced dependence
of the accuracy on the equivalent width itself. The relation found is
best described by

\begin{equation}
\begin{array}{c}

      e = 0.035 + 0.088 w

\end{array}
\end{equation}

where $e$ is the average accuracy with which a line of equivalent width $w$ is
determined. Units are Angstroms.
A complete list of the lines is given in Table~2, with their
wavelength taken from the Multiplet Table by Ch. Moore (1972) as well as
their identification. Furthermore, for each line, an inner region and two outer regions
were selected. The two outer regions, one on each side of the line -each
al least several Angstroms wide- were used to determine a "continuum" or
"pseudo-continuum", while the integration of the inner region yielded the 
equivalent width. Also it is indicated whether they were
measured in the B- or in the V-spectra or in both. All this information is
shown in Table~2.
For two lines the relation between the equivalent widths measured
in the B- and the V-spectra is shown in Figure~3. Likewise the relation
between the intrinsic colors $(B-V)_0$ and the equivalent widths for 
H-$\delta$ and HeI(4143) is shown in
Figure~4. In these plots the size of the symbol represents the luminosity
of the star in the sense that the larger the symbol the more luminous
the star. The plots demonstrate the known fact that the He-lines
reach their maximum equivalent width around the spectral type B2, while the
equivalent width of the hydrogen lines increases all the way through the B-class.

\section{Analysis}

The principal purpose of this work is to find means by which physical
parameters, namely the absolute magnitude $Mv$ and the intrinsic
color $(B-V)_0$, can be predicted from the equivalent widths of
absorption lines. The external data used are the spectral types in
the MK-system, the apparent V-magnitudes, and the observed $(B-V)$-
colors given in the Bright Star catalogue, and the parallaxes and
parallax errors in the Hipparcos Catalogue.
Intrinsic colors $(B-V)_0$ were deduced from the MK-types. The
respective relations may be found for instance in tables given by
Allen (1973). These colors can be compared with the observed colors
and a reddening  effect can be found. If interpreted as due to
absorption by interstellar or circumstellar material the effect on
the apparent V-magnitude can be estimated. For this purpose we use
the usually adopted relation

\begin{equation}
\begin{array}{c}

                    Av = 3.0 E((B-V))

\end{array}
\end{equation}

where $Av$ is the absorption in the V-band, and $E((B-V))$ the reddening
of the $(B-V)$-color. This correction was applied when the color excess
is greater than 0.03 magnitudes. Using the parallax given in the
Hipparcos Parallax Catalogue the corrected magnitudes were converted
into absolute magnitudes $Mv$.
For the relation between the physical parameters $Mv$  or
$(B-V)_0$ and the equivalent widths
we adopt second order polynomials with two or three independent
variables, the latter being the equivalent widths of two or three
absorption lines. Thus the equation for the absolute magnitude with
three lines has the form

\begin{equation}
\begin{array}{cll}


Mv = & a_{000} +
            a_{100} w_1 +
            a_{010} w_2 +
            a_{001} w_3 \\
          &
            + a_{200} {w_1}^2 
            + a_{110} w_1 w_2 + 
            a_{101} w_1 w_3 \\
          &
            + a_{020} {w_2}^2 
            + a_{011} w_2 w_3 +
            a_{002} {w_3}^2

\end{array}
\end{equation}

where $w1$,$w2$, and $w3$ are the equivalent widths of the three respective
absorption lines. The coefficients $a_{ijk}$ have to be determined by least
squares, forming equation (3) for all spectra in which values for the
three different equivalent widths were obtained. With 3 lines out
of 12 a total of 220 combinations can be made up. With two lines a
total of 66 combinations can be made up. Equation (3) can also be used
for the intrinsic colors, introducing these instead of the absolute
magnitudes. For the three-line combinations 10 coefficients have to
be determined by least squares. If only two lines are used (skipping
in equation (3) all terms containing $w_3$) the number of unknowns is 6.
Likewise omitting the square and mixed terms the expression for the
linear dependence of $Mv$ on two or three lines is obtained.
Replacing in all equations $Mv$ by $(B-V)_0$ expressions are found
which relate the intrinsic colors to the equivalent widths.
We have also tested solutions based on four absorption lines,
without obtaining a significant improvement with respect to the previous
solutions.

We should point out here that all spectra entered the solution
with the same weight. It would be possible to assign weight according
to the respective relative parallax errors, i.e. the parallax error
divided by the parallax itself, which in our selection is limited to
a maximum of 20\%. This idea was discarded in view of the biased
selection of stars in the Hipparcos Catalogue which may favor stars
of certain absolute magnitudes, and of the fact that in general low
luminosity stars have more accurate parallaxes due to their proximity.

\section{Results for the absolute magnitudes}

Two- and three-line solutions were calculated for all possible line
combinations. For each combination the average residual $r_{av}$ (Hipparcos
absolute magnitude minus polynomial absolute magnitude) was calculated.
Outliers with residuals larger than $2.5 r_{av}$ were eliminated and the
solution was repeated. For both the two-line and the three-line
solution those with the smallest $r_{av}$ were selected. The respective
combinations for two and three lines and their $r_{av}$-value are given 
in Table~3 and Table~4. The full
information with the corresponding coefficients is found in Table~3a
and Table~4a
available on our website (www.cida.ve/\verb+~+stock/paper2). We also tried the linear
dependence of the absolute magnitude on the equivalent widths of two
or three lines. The respective data are also given in Table~3, Table~4\
Table~3a and Table~4a.
As may be seen in these tables average residuals of 0.40 magnitudes can be 
obtained. For the best combination of the lineal solutions with three lines 
we also give the respective coefficients in Table~5.

The importance of the line 12 (HeI 5047.736) is evident. Since
this line is present only in the V-spectra which are considerably
less numerous than the B-spectra we have also calculated solutions
based entirely on the 11 lines in the B-spectra. The respective
results are also given in Table~3-4 and Table~3a-4a. Comparing the data
for the solutions with 11 and with 12 lines the importance of the
inclusion of the V-spectra  is clearly demonstrated.
For most of the stars on our list comments are given in the
Bright Star Catalogue, some actually rather extensive. We have looked
at the comments for the outliers eliminated in the calculation of the
coefficients. They do have comments, but these are shared with other
stars which were not found to be outliers. It appears that we would
need more stars if we were to determine whether the "outlier" condition
can be predicted on the basis of measurements of the equivalent widths
of absorption lines only. We should point out here that only two
supergiants are included in our sample of stars, one a B8.Ia, the other
a O9.5Ib. Both turned out to be outliers.
Consistently solutions based on three lines gave better results
than those based on two lines. The differences between the linear and
the second order solutions, however, indicate no clear advantage of
one or the other.

\section{Results for the intrinsic colors}

We have already indicated that equation (3) can readily be modified
to be applied to the intrinsic colors as function of the equivalent
widths of two or three lines in a linear or second order polynomial.
The data for the combinations which gave the smallest average
residual are given in Table~6-7 and Table~6a-7a, the latter two available
on our website (www.cida.ve/\verb+~+stock/paper2). As may be seen in these
tables, average residuals of 0.016 magnitudes can be obtained.
Just as in the case of the absolute magnitudes a significant
improvement is obtained with the three-line dependence as compared
to the two-line dependence. On the other hand, linear or second
order solutions do not consistently favor one or the other. The lines
10 (HeI 4921.929) and 12 (HeI 5047.736) occur most frequently
among the ten best solutions.

\section{Conclusions}

For the recovery of the physical stellar parameters $Mv$ and $(B-V)_0$
from the equivalent widths of absorption lines we have tested
linear and second order polynomials with two and with three lines
as independent variables, making use of all possible line
combinations. Significant improvement was obtained by switching
from a two-line dependence to a three-line dependence, but no
consistent improvement was found by switching from linear to second
order polynomials for both the absolute magnitudes and the intrinsic
colors.
Table~8 shows how often the different lines were used in the
best solutions for $Mv$ and for $(B-V)_0$ colors. Taking into account 
that line 12 only occurs
in the V-spectra which are far less numerous than the B-spectra the
lines 4, 9, 11, and 12 are the most important lines for the recovery
of the absolute magnitudes.
For intrinsic colors, again one has to allow for the fact that
line 12 is far less observed than all the other lines. The lines
3, 8, 10, and 12 are the most important ones for the recovery of the
intrinsic colors. The usefulness of the hidrogen lines H-$\beta$ (9) and
H-$\gamma$ (4) has long been known. The sensivity of the green helium lines
5015.7 (11) and 5047.7 (12) for classification purposes was not widely
known because their are not available on spectra taken for MK classification.

\acknowledgments

This research has been partially supported by
CONDES of the Universidad del Zulia and by CONICIT
from Venezuela.

\onecolumn

\newpage

\begin{center}
\tablecaption{List of observed stars}
\tablefirsthead{
\hline\hline
HD number & V & (B-V) & Parallax (mas) & Parallax error (mas)  & Spectral type & B-sp & V-sp \\
\hline }
\tablehead{%
\multicolumn{8}{l}{\small\sl continued from previous page} \\
\hline
HD number & V & (B-V) & Parallax (mas) & Parallax error (mas)  & Spectral type & B-sp & V-sp \\
\hline }
\tabletail{%
\hline
\multicolumn{8}{r}{\small\sl continued on next page...} \\ }
\tablelasttail{
\hline}
\begin{supertabular}{cccccccc}


\hline

 224990 &     5.04 &    -0.15 &     6.40 &     0.87 &   B4 V          &  * &    \\
    886 &     2.83 &    -0.19 &     9.79 &     0.81 &   B2 IV         &  * &    \\
   3369 &     4.34 &    -0.12 &     4.97 &     0.82 &   B5 V          &  * &    \\
   5737 &     4.30 &    -0.15 &     4.85 &     0.84 &   B7 IIIp       &  * &    \\
   7374 &     5.97 &    -0.08 &     6.52 &     0.79 &   B8 III        &  * &    \\
  14951 &     5.48 &    -0.10 &     5.41 &     1.04 &   B7 IV         &  * &    \\
  16582 &     4.08 &    -0.21 &     5.04 &     0.83 &   B2 IV         &  * &    \\
  17081 &     4.24 &    -0.12 &     7.40 &     0.85 &   B7 IV         &  * &    \\
  18604 &     4.71 &    -0.11 &     7.69 &     0.76 &   B6 III        &  * &  * \\
  19356 &     2.09 &     0.00 &    35.14 &     0.90 &   B8 V          &  * &  * \\
  22203 &     4.26 &    -0.11 &    11.02 &     0.75 &   B9 V          &  * &    \\
  23227 &     4.99 &    -0.16 &     4.45 &     0.62 &   B5 III        &  * &    \\
  23302 &     3.72 &    -0.10 &     8.80 &     0.89 &   B6 III        &  * &    \\
  23338 &     4.30 &    -0.11 &     8.75 &     1.08 &   B6 V          &  * &    \\
  23466 &     5.34 &    -0.10 &     5.54 &     0.80 &   B3 V          &  * &    \\
  23630 &     2.85 &    -0.09 &     8.87 &     0.99 &   B7 III        &  * &  * \\
  23850 &     3.62 &    -0.07 &     8.57 &     1.03 &   B8 III        &  * &  * \\
  23277 &     5.40 &     0.10 &    10.02 &     0.54 &   A2 m          &  * &    \\
  24587 &     4.64 &    -0.14 &     8.46 &     0.75 &   B5 V          &  * &    \\
  24760 &     2.90 &    -0.20 &     6.06 &     0.82 &   B0.5 V        &  * &    \\
  25340 &     5.28 &    -0.13 &     7.20 &     0.83 &   B5 V          &  * &    \\
  25330 &     5.67 &     0.00 &     5.77 &     0.78 &   B5 V          &  * &    \\
  26326 &     5.45 &    -0.15 &     4.49 &     0.78 &   B5 IV         &  * &  * \\
  26912 &     4.27 &    -0.05 &     7.50 &     1.13 &   B3 IV         &  * &  * \\
  28375 &     5.53 &    -0.10 &     8.47 &     1.11 &   B3 V          &  * &    \\
  29248 &     3.93 &    -0.21 &     5.56 &     0.88 &   B2 III sb     &  * &    \\
  29763 &     4.27 &    -0.11 &     8.14 &     0.78 &   B3 V          &  * &  * \\
  30211 &     4.01 &    -0.15 &     6.13 &     1.03 &   B5 IV         &  * &  * \\
  33802 &     4.45 &    -0.10 &    13.53 &     0.69 &   B8 V          &  * &  * \\
  34085 &     0.18 &    -0.03 &     4.22 &     0.81 &   B8 Ia         &  * &  * \\
  34503 &     3.59 &    -0.12 &     5.88 &     0.77 &   B5 III        &  * &  * \\
  35468 &     1.64 &    -0.22 &    13.42 &     0.98 &   B2 III        &  * &  * \\
  35497 &     1.65 &    -0.13 &    24.89 &     0.88 &   B7 III        &  * &  * \\
  36267 &     4.20 &    -0.14 &    11.30 &     1.01 &   B5 V          &  * &  * \\
  37742 &     1.74 &    -0.20 &     3.99 &     0.79 &   O9.5 Ib sb    &    &  * \\
  41534 &     5.65 &    -0.19 &     3.01 &     0.57 &   B2 V          &  * &    \\
  41753 &     4.42 &    -0.16 &     6.10 &     0.88 &   B3 IV         &  * &  * \\
  42560 &     4.45 &    -0.18 &     5.14 &     0.78 &   B3 IV         &  * &  * \\
  43157 &     5.83 &    -0.16 &     5.28 &     0.83 &   B5 V          &  * &    \\
  43153 &     5.34 &    -0.10 &     6.80 &     0.93 &   B7 V          &  * &  * \\
  43955 &     5.51 &    -0.16 &     3.28 &     0.60 &   B2/B3 V       &  * &    \\
  44743 &     1.98 &    -0.24 &     6.53 &     0.66 &   B1 II/III     &  * &  * \\
  45813 &     4.47 &    -0.17 &     8.03 &     0.58 &   B4 V          &  * &  * \\
  45542 &     4.13 &    -0.12 &     6.49 &     1.06 &   B6 III        &  * &  * \\
  46487 &     5.09 &    -0.13 &     6.08 &     0.79 &   B5 Vn         &  * &  * \\
  46936 &     5.62 &    -0.09 &     6.70 &     0.58 &   B9 V          &  * &  * \\
  47100 &     5.34 &    -0.08 &     4.30 &     0.76 &   B8 III        &  * &    \\
  49643 &     5.75 &    -0.10 &     5.84 &     0.77 &   B8 IIIn       &  * &    \\
  52089 &     1.50 &    -0.21 &     7.57 &     0.57 &   B2 II         &  * &  * \\
  52670 &     5.64 &    -0.17 &     3.17 &     0.59 &   B2/B3 III/IV  &  * &  * \\
  53244 &     4.11 &    -0.11 &     8.11 &     0.63 &   B8 II         &  * &  * \\
  56342 &     5.36 &    -0.16 &     5.05 &     0.57 &   B2 V          &  * &    \\
  57821 &     4.94 &    -0.04 &     6.31 &     0.69 &   B5 II/III     &    &  * \\
  58715 &     2.89 &    -0.10 &    19.16 &     0.85 &   B8 Vvar       &  * &    \\
  59550 &     5.78 &    -0.19 &     2.85 &     0.56 &   B2 IV         &    &  * \\
  60863 &     4.65 &    -0.11 &    14.50 &     0.62 &   B8 V          &  * &  * \\
  61429 &     4.69 &    -0.10 &     5.92 &     0.72 &   B8 IV         &  * &  * \\
  61555 &     3.80 &    -0.16 &     7.18 &     1.06 &   B5 IV         &  * &    \\
  63975 &     5.12 &    -0.12 &     7.76 &     1.02 &   B8 II         &  * &  * \\
  67797 &     4.40 &    -0.16 &     6.90 &     0.69 &   B5 V          &  * &  * \\
  78316 &     5.23 &    -0.09 &     6.74 &     0.91 &   B8 IIImnp     &  * &  * \\
  83754 &     5.07 &    -0.15 &     6.33 &     0.91 &   B4 IV/V       &    &  * \\
  87901 &     1.36 &    -0.09 &    42.09 &     0.79 &   B7 V          &  * &  * \\
  90994 &     5.08 &    -0.14 &     9.46 &     1.16 &   B6 V          &  * &  * \\
 106625 &     2.58 &    -0.11 &    19.78 &     0.81 &   B8 III        &    &  * \\
 107348 &     5.20 &    -0.09 &     8.47 &     0.78 &   B8 V          &  * &  * \\
 116658 &     0.98 &    -0.23 &    12.44 &     0.86 &   B1 V          &  * &  * \\
 120315 &     1.85 &    -0.10 &    32.39 &     0.74 &   B3 V sb       &  * &  * \\
 120709 &     4.32 &    -0.15 &    10.96 &     0.88 &   B5           &  * &  * \\
 120955 &     4.75 &    -0.11 &     4.87 &     0.71 &   B4 IV         &  * &  * \\
 121847 &     5.20 &    -0.09 &     9.61 &     0.69 &   B8 V          &  * &    \\
 126769 &     4.97 &    -0.07 &     7.85 &     0.80 &   B7/B8 V       &  * &  * \\
 132955 &     5.45 &    -0.13 &     9.32 &     0.84 &   B3 V          &  * &    \\
 135742 &     2.61 &    -0.07 &    20.38 &     0.87 &   B8 V          &  * &  * \\
 136298 &     3.22 &    -0.23 &     6.39 &     0.86 &   B1.5 IV       &  * &    \\
 138485 &     5.53 &    -0.15 &     4.24 &     0.84 &   B3 V          &  * &    \\
 138749 &     4.14 &    -0.13 &    10.49 &     0.66 &   B6 Vnn        &  * &    \\
 138764 &     5.16 &    -0.09 &     9.30 &     0.86 &   B6 IV         &  * &    \\
 139365 &     3.66 &    -0.18 &     7.33 &     1.01 &   B2.5 V        &  * &    \\
 142883 &     5.84 &     0.01 &     7.16 &     0.87 &   B3 V          &  * &    \\
 143275 &     2.29 &    -0.12 &     8.12 &     0.88 &   B0.2 IV       &  * &    \\
 147394 &     3.91 &    -0.15 &    10.37 &     0.53 &   B5 IV         &  * &    \\
 148605 &     4.79 &    -0.12 &     8.30 &     0.84 &   B3 V          &  * &    \\
 148703 &     4.24 &    -0.17 &     4.37 &     0.80 &   B2 III-IV     &  * &    \\
 149438 &     2.82 &    -0.21 &     7.59 &     0.78 &   B0 V          &  * &    \\
 149757 &     2.54 &     0.04 &     7.12 &     0.71 &   O9.5 V        &  * &    \\
 154204 &     6.29 &    -0.04 &     8.21 &     0.76 &   B7 IV/V       &  * &    \\
 158408 &     2.70 &    -0.18 &     6.29 &     0.81 &   B2 IV         &  * &    \\
 158926 &     1.62 &    -0.23 &     4.64 &     0.90 &   B1.5 IV+...   &  * &    \\
 160762 &     3.82 &    -0.18 &     6.58 &     0.56 &   B3 V sb       &  * &    \\
 160578 &     2.39 &    -0.17 &     7.03 &     0.73 &   B1.5 III      &  * &    \\
 172910 &     4.86 &    -0.17 &     7.23 &     1.09 &   B2 V          &  * &    \\
 173300 &     3.17 &    -0.11 &    14.14 &     0.88 &   B8.5 III      &  * &    \\
 175191 &     2.05 &    -0.13 &    14.54 &     0.88 &   B2.5 V        &  * &    \\
 176162 &     5.51 &    -0.04 &     6.34 &     0.80 &   B4 V          &  * &    \\
 180163 &     4.43 &    -0.15 &     3.13 &     0.51 &   B2.5 IV       &  * &    \\
 180554 &     4.76 &    -0.06 &     3.58 &     0.60 &   B4 IV         &  * &    \\
 182255 &     5.22 &    -0.12 &     8.10 &     0.68 &   B6 III        &  * &    \\
 182568 &     4.99 &    -0.12 &     4.21 &     0.58 &   B3 IV         &  * &    \\
 184171 &     4.74 &    -0.15 &     5.20 &     0.55 &   B3 IV         &  * &    \\
 184930 &     4.36 &    -0.08 &    10.61 &     0.94 &   B5 III        &  * &    \\
 186500 &     5.51 &     0.02 &     5.98 &     1.00 &   B8 III        &  * &    \\
 189103 &     4.37 &    -0.15 &     5.28 &     0.90 &   B2.5 IV       &  * &    \\
 189944 &     5.88 &    -0.13 &     4.85 &     0.66 &   B4 V          &  * &    \\
 190993 &     5.08 &    -0.16 &     6.68 &     0.71 &   B3 V          &  * &    \\
 196740 &     5.06 &    -0.13 &     6.61 &     0.66 &   B5 IV         &  * &    \\
 202671 &     5.40 &    -0.12 &     5.63 &     0.95 &   B5 II/III     &  * &    \\
 207330 &     4.23 &    -0.12 &     2.82 &     0.52 &   B3 III        &  * &    \\
 207971 &     3.00 &    -0.08 &    16.07 &     0.77 &   B8 III        &  * &    \\
 209409 &     4.74 &    -0.10 &     8.56 &     0.81 &   B7 IVe        &  * &    \\
 210424 &     5.43 &    -0.12 &     5.81 &     0.75 &   B5 III        &  * &    \\
 210934 &     5.45 &    -0.12 &     6.42 &     0.85 &   B7 V          &  * &    \\
 214748 &     4.18 &    -0.10 &     4.38 &     0.87 &   B8 V          &  * &    \\
 214923 &     3.41 &    -0.09 &    15.64 &     0.75 &   B8.5 V        &  * &    \\
 216831 &     5.73 &    -0.05 &     3.90 &     0.70 &   B7 III        &  * &    \\
 219688 &     4.41 &    -0.14 &    10.13 &     1.04 &   B5 Vn         &  * &    \\

\hline
\hline
\end{supertabular}
\end{center}

\begin{table}
\caption{List of selected Lines}
\begin{center}
\begin{tabular}{ccccccccc}\hline\hline

Line  & \multicolumn{2}{c}{Cont1(\AA)} & \multicolumn{2}{c}{Line(\AA)} & \multicolumn{2}{c}{Cont2(\AA)} & $\lambda$ (\AA)  &  Id. \\

\hline

1 B & 4006.2 & 4020.3 & 4022.1 & 4032.7 & 4036.2 & 4050.3 & 4026.4 & HeI \\
2 B & 4067.9 & 4082.1 & 4085.6 & 4119.1 & 4124.4 & 4142.0 & 4101.7 & H$\delta$ \\
3 B & 4122.6 & 4140.2 & 4142.0 & 4156.1 & 4157.9 & 4168.4 & 4143.8 & HeI \\
4 B & 4298.9 & 4313.0 & 4316.5 & 4357.1 & 4362.4 & 4374.7 & 4340.5 & H$\gamma$ \\
5 B & 4364.1 & 4378.2 & 4381.8 & 4392.3 & 4397.6 & 4410.0 & 4387.9 & HeI \\
6 B & 4439.9 & 4462.9 & 4454.0 & 4477.0 & 4478.7 & 4487.5 & 4471.7 & HeI \\
7 B & 4475.2 & 4477.0 & 4478.7 & 4482.3 & 4485.8 & 4494.6 & 4481.0 & MgII \\
8 B & 4686.8 & 4702.6 & 4709.7 & 4716.7 & 4720.3 & 4729.1 & 4713.4 & HeI \\
9 B-V & 4820.8 & 4833.1 & 4836.6 & 4868.0 & 4889.5 & 4905.4 & 4861.3 & H$\beta$ \\
10 B-V & 4900.1 & 4912.4 & 4916.0 & 4926.5 & 4931.8 & 4945.9 & 4921.3 & HeI \\
11 B-V & 5005.9 & 5011.2 & 5014.7 & 5020.0 & 5023.5 & 5032.3 & 5015.7 & HeI \\
12 V & 5101.1 & 5112.4 & 5116.9 & 5129.3 & 5136.3 & 5148.7 & 5047.7 & HeI \\

\hline
\end{tabular}
\end{center}
\end{table}

\begin{table}
\caption{Best combinations for the determination of Absolute Magnitudes based in Two lines}
\begin{center}
\begin{tabular}{c}
Linear Solutions
\end{tabular} \\
\begin{tabular}{cccccccccccc}\hline\hline
 & \multicolumn{3}{c}{B-spectra} & & & & & \multicolumn{3}{c}{B- and  V-spectra} &  \\
\hline

L1 & L2 & $r_{av}$ & N$_u$ & N$_e$ &  &  & L1 & L2 & $r_{av}$ & N$_u$ & N$_e$  \\

\hline

  2 &  4 &   0.517 &   54  &   5  &  &  &  2 &  4 &   0.517 &   54 &    5 \\
  2 &  9 &   0.504 &   54  &   5  &  &  &  2 &  9 &   0.504 &   54 &    5 \\
  3 &  9 &   0.529 &   91  &   9  &  &  &  3 &  9 &   0.529 &   91 &    9 \\
  4 &  7 &   0.553 &  102  &   9  &  &  &  4 &  7 &   0.553 &  102 &    9 \\
  4 & 11 &   0.496 &   74  &   7  &  &  &  4 & 11 &   0.496 &   74 &    7 \\
  5 &  9 &   0.588 &  103  &   8  &  &  &  4 & 12 &   0.381 &   34 &    5 \\
  6 &  9 &   0.581 &  102  &   9  &  &  &  7 &  9 &   0.556 &  102 &    9 \\
  7 &  9 &   0.556 &  102  &   9  &  &  &  9 & 10 &   0.568 &  106 &   10 \\
  9 & 10 &   0.568 &  106  &  10  &  &  &  9 & 11 &   0.563 &   82 &    4 \\
  9 & 11 &   0.563 &   82  &   4  &  &  &  9 & 12 &   0.482 &   41 &    3 \\

\hline
\end{tabular} \\

\begin{tabular}{c}
Quadratic Solutions
\end{tabular} \\
\begin{tabular}{cccccccccccc}\hline
 & \multicolumn{3}{c}{B-spectra} & & & & & \multicolumn{3}{c}{B- and  V-spectra} &  \\
\hline
L1 & L2 & $r_{av}$ & N$_u$ & N$_e$ &  &  & L1 & L2 & $r_{av}$ & N$_u$ & N$_e$  \\
\hline

  2 &  9 &   0.460 &   54  &   5  &  &  &  2 &  9 &   0.460 &   54 &    5 \\
  3 &  9 &   0.550 &   93  &   7  &  &  &  3 &  9 &   0.550 &   93 &    7 \\
  4 &  7 &   0.604 &  104  &   7  &  &  &  4 &  8 &   0.578 &  102 &    9 \\
  4 &  8 &   0.578 &  102  &   9  &  &  &  4 &  9 &   0.561 &  103 &    8 \\
  4 &  9 &   0.561 &  103  &   8  &  &  &  4 & 11 &   0.513 &   74 &    7 \\
  4 & 11 &   0.513 &   74  &   7  &  &  &  7 &  9 &   0.572 &  104 &    7 \\
  7 &  9 &   0.572 &  104  &   7  &  &  &  8 &  9 &   0.536 &  102 &    9 \\
  8 &  9 &   0.536 &  102  &   9  &  &  &  9 & 10 &   0.567 &  108 &    8 \\
  9 & 10 &   0.567 &  108  &   8  &  &  &  9 & 11 &   0.586 &   84 &    2 \\
  9 & 11 &   0.586 &   84  &   2  &  &  &  9 & 12 &   0.412 &   41 &    3 \\

\hline
\multicolumn{12}{c}{\small\sl N$_u$:used stars, N$_e$:eliminated stars} \\ 

\end{tabular}
\end{center}
\end{table}

\begin{table}
\caption{Best combinations for the determination of Absolute Magnitudes based in Three lines}
\begin{center}
\begin{tabular}{c}
Linear Solutions
\end{tabular} \\
\begin{tabular}{cccccccccccccc}\hline\hline

 & \multicolumn{4}{c}{B-spectra} & & & & & \multicolumn{4}{c}{B- and  V-spectra} &  \\
\hline

L1 & L2 & L3 & $r_{av}$ & N$_u$ & N$_e$ &  &  & L1 & L2 & L3 & $r_{av}$ & N$_u$ & N$_e$  \\

\hline

  1 &  2 &  9 &  0.482 &  53 &  5 & & &  1 &  4 & 11 &  0.421 &  26 &  2 \\
  1 &  4 & 11 &  0.421 &  26 &  2 & & &  1 &  4 & 12 &  0.387 &  26 &  2 \\
  1 &  9 & 11 &  0.333 &  26 &  2 & & &  1 &  9 & 11 &  0.333 &  26 &  2 \\
  2 &  4 & 11 &  0.451 &  28 &  1 & & &  1 &  9 & 12 &  0.326 &  26 &  2 \\
  2 &  9 & 10 &  0.498 &  54 &  5 & & &  2 &  4 & 11 &  0.451 &  28 &  1 \\
  3 &  4 & 11 &  0.464 &  64 &  6 & & &  2 &  4 & 12 &  0.440 &  27 &  1 \\
  3 &  9 & 11 &  0.475 &  65 &  5 & & &  2 &  9 & 12 &  0.384 &  27 &  1 \\
  4 &  5 & 11 &  0.500 &  74 &  7 & & &  3 &  4 & 11 &  0.464 &  64 &  6 \\
  4 &  8 & 11 &  0.499 &  74 &  7 & & &  4 &  8 & 12 &  0.424 &  35 &  4 \\
  4 & 10 & 11 &  0.486 &  74 &  7 & & &  4 & 10 & 12 &  0.387 &  34 &  5 \\

\hline
\end{tabular} \\

\begin{tabular}{c}
Quadratic Solutions
\end{tabular} \\
\begin{tabular}{cccccccccccccc}\hline
 & \multicolumn{4}{c}{B-spectra} & & & & & \multicolumn{4}{c}{B- and  V-spectra} &  \\
\hline
L1 & L2 & L3 & $r_{av}$ & N$_u$ & N$_e$ &  &  & L1 & L2 & L3 & $r_{av}$ & N$_u$ & N$_e$  \\
\hline

  2 &   4 &  9 &   0.465 &   56 & 3 & &    & 1 &  9 & 12 &   0.395 &   27 & 1 \\
  2 &   5 &  9 &   0.489 &   55 & 4 & &    & 2 &  6 &  9 &   0.443 &   54 & 5 \\
  2 &   6 &  9 &   0.443 &   54 & 5 & &    & 3 &  9 & 11 &   0.447 &   66 & 4 \\
  2 &   7 &  9 &   0.484 &   55 & 4 & &    & 3 &  9 & 12 &   0.334 &   35 & 4 \\
  2 &   9 & 10 &   0.483 &   55 & 4 & &    & 4 &  6 & 11 &   0.456 &   75 & 6 \\
  3 &   9 & 11 &   0.447 &   66 & 4 & &    & 4 &  8 & 12 &   0.457 &   38 & 1 \\
  4 &   6 & 11 &   0.456 &   75 & 6 & &    & 4 &  9 & 12 &   0.417 &   37 & 2 \\
  4 &   8 & 11 &   0.473 &   75 & 6 & &    & 7 &  9 & 12 &   0.454 &   38 & 1 \\
  4 &   9 & 11 &   0.471 &   76 & 5 & &    & 8 &  9 & 12 &   0.440 &   38 & 1 \\
  9 &  10 & 11 &   0.480 &   81 & 5 & &    & 9 & 10 & 12 &   0.342 &   40 & 4 \\

\hline
\multicolumn{12}{c}{\small\sl N$_u$:used stars, N$_e$:eliminated stars} \\ 

\end{tabular}
\end{center}
\end{table}

\begin{table}
\caption{Coefficients for the best linear combinations of Absolute Magnitudes based in Three lines}
\begin{center}
\begin{tabular}{c}
Blue Spectra
\end{tabular} \\
\begin{tabular}{ccccccc}\hline\hline

L1 & L2 & L3 & $a_{000}$ & $a_{100}$ & $a_{010}$ & $a_{001}$ \\

\hline

  1 &  2 &  9 &  -5.1881 &   0.1720 &  -0.3517 &   0.9100 \\
  1 &  4 & 11 &  -6.1041 &  -1.1267 &   0.8186 &   4.2092 \\
  1 &  9 & 11 &  -5.3635 &  -0.2788 &   0.6886 &   0.0740 \\
  2 &  4 & 11 &  -6.2617 &  -0.5080 &   1.2510 &   1.1302 \\
  2 &  9 & 10 &  -4.5249 &  -0.3179 &   0.8207 &  -0.5408 \\
  3 &  4 & 11 &  -5.4876 &   0.5196 &   0.7536 &  -1.4535 \\
  3 &  9 & 11 &  -5.0903 &   0.5743 &   0.6669 &  -1.4267 \\
  4 &  5 & 11 &  -5.0534 &   0.6887 &   0.0973 &  -1.8178 \\
  4 &  8 & 11 &  -4.9882 &   0.6807 &  -0.3788 &  -1.4161 \\
  4 & 10 & 11 &  -5.4992 &   0.7414 &   0.8885 &  -2.7578 \\

\hline
\end{tabular} \\

\begin{tabular}{c}
Blue and Visual Spectra
\end{tabular} \\
\begin{tabular}{ccccccc}\hline

L1 & L2 & L3 & $a_{000}$ & $a_{100}$ & $a_{010}$ & $a_{001}$ \\

\hline

  1 &  4 & 11 &  -6.1041 &  -1.1267 &   0.8186 &   4.2092 \\
  1 &  4 & 12 &  -5.6824 &  -0.1997 &   0.7742 &   5.1345 \\
  1 &  9 & 11 &  -5.3635 &  -0.2788 &   0.6886 &   0.0740 \\
  1 &  9 & 12 &  -5.3137 &  -0.2258 &   0.6787 &   1.6170 \\
  2 &  4 & 11 &  -6.2617 &  -0.5080 &   1.2510 &   1.1302 \\
  2 &  4 & 12 &  -5.8773 &  -0.2922 &   1.0317 &   4.0268 \\
  2 &  9 & 12 &  -5.4125 &  -0.1425 &   0.7869 &   0.9973 \\
  3 &  4 & 11 &  -5.4876 &   0.5196 &   0.7536 &  -1.4535 \\
  4 &  8 & 12 &  -5.4024 &   0.7490 &  -1.1068 &   5.4555 \\
  4 & 10 & 12 &  -5.9997 &   0.8121 &   0.1954 &   5.1382 \\

\hline
\end{tabular}
\end{center}
\end{table}

\begin{table}
\caption{Best combinations for the determination of $(B-V)_0$ colors based in Two lines}
\begin{center}
\begin{tabular}{c}
Linear Solutions
\end{tabular} \\
\begin{tabular}{cccccccccccc}\hline\hline
 & \multicolumn{3}{c}{B-spectra} & & & & & \multicolumn{3}{c}{B- and  V-spectra} &  \\
\hline

L1 & L2 & $r_{av}$ & N$_u$ & N$_e$ &  &  & L1 & L2 & $r_{av}$ & N$_u$ & N$_e$  \\

\hline

  1 & 10 &  0.020 &   55 &   3 & & &  1 & 10 &  0.020 &  55 &   3 \\
  3 &  5 &  0.018 &   87 &  13 & & &  3 &  5 &  0.018 &  87 &  13 \\
  3 &  6 &  0.019 &   91 &   9 & & &  3 &  6 &  0.019 &  91 &   9 \\
  3 & 10 &  0.017 &   88 &  12 & & &  3 & 10 &  0.017 &  88 &  12 \\
  4 & 10 &  0.020 &  102 &   9 & & &  5 & 10 &  0.020 &  99 &  12 \\
  5 & 10 &  0.020 &   99 &  12 & & &  5 & 12 &  0.018 &  36 &   3 \\
  6 & 10 &  0.021 &  101 &  10 & & &  7 & 10 &  0.020 &  98 &  13 \\
  7 & 10 &  0.020 &   98 &  13 & & &  8 & 12 &  0.019 &  37 &   2 \\
  9 & 10 &  0.020 &  108 &   8 & & & 10 & 11 &  0.019 &  74 &  12 \\
 10 & 11 &  0.019 &   74 &  12 & & & 10 & 12 &  0.017 &  41 &   3 \\

\hline
\end{tabular} \\

\begin{tabular}{c}
Quadratic Solutions
\end{tabular} \\
\begin{tabular}{cccccccccccc}\hline
 & \multicolumn{3}{c}{B-spectra} & & & & & \multicolumn{3}{c}{B- and  V-spectra} &  \\
\hline
L1 & L2 & $r_{av}$ & N$_u$ & N$_e$ &  &  & L1 & L2 & $r_{av}$ & N$_u$ & N$_e$  \\
\hline

  1 & 10 &  0.021 &  55 &  3 & & &  1 & 10 &  0.021 &  55 &  3 \\
  2 &  6 &  0.022 &  58 &  1 & & &  2 &  6 &  0.022 &  58 &  1 \\
  2 & 10 &  0.022 &  58 &  1 & & &  3 & 10 &  0.021 &  92 &  8 \\
  3 & 10 &  0.021 &  92 &  8 & & &  4 &  6 &  0.021 & 103 &  8 \\
  4 &  5 &  0.023 & 103 &  8 & & &  4 & 10 &  0.021 & 102 &  9 \\
  4 &  6 &  0.021 & 103 &  8 & & &  6 &  9 &  0.022 & 105 &  6 \\
  4 & 10 &  0.021 & 102 &  9 & & &  8 & 10 &  0.021 &  99 & 12 \\
  6 &  9 &  0.022 & 105 &  6 & & &  8 & 12 &  0.020 &  38 &  1 \\
  8 & 10 &  0.021 &  99 & 12 & & &  9 & 10 &  0.020 & 109 &  7 \\
  9 & 10 &  0.020 & 109 &  7 & & & 10 & 12 &  0.019 &  41 &  3 \\

\hline
\multicolumn{12}{c}{\small\sl N$_u$:used stars, N$_e$:eliminated stars} \\ 

\end{tabular}
\end{center}
\end{table}

\begin{table}
\caption{Best combinations for the determination of $(B-V)_0$ colors based in Three lines}
\begin{center}
\begin{tabular}{c}
Linear Solutions
\end{tabular} \\
\begin{tabular}{cccccccccccccc}\hline\hline

 & \multicolumn{4}{c}{B-spectra} & & & & & \multicolumn{4}{c}{B- and  V-spectra} &  \\
\hline

L1 & L2 & L3 & $r_{av}$ & N$_u$ & N$_e$ &  &  & L1 & L2 & L3 & $r_{av}$ & N$_u$ & N$_e$  \\

\hline

  1 &  3 &  6 &  0.018 &  53 &  5 & & & 1 &  8 & 12 &  0.017 &  27 &  1 \\
  1 &  3 & 10 &  0.018 &  54 &  4 & & & 2 &  8 & 11 &  0.014 &  27 &  2 \\
  1 &  8 & 11 &  0.017 &  27 &  1 & & & 2 &  8 & 12 &  0.016 &  27 &  1 \\
  2 &  8 & 11 &  0.014 &  27 &  2 & & & 3 &  5 & 12 &  0.016 &  36 &  3 \\
  3 &  5 &  7 &  0.018 &  87 & 13 & & & 3 &  6 & 12 &  0.016 &  36 &  3 \\
  3 &  5 & 10 &  0.019 &  91 &  9 & & & 3 & 10 & 12 &  0.016 &  37 &  2 \\
  3 &  6 & 10 &  0.019 &  92 &  8 & & & 4 &  8 & 12 &  0.013 &  35 &  4 \\
  3 &  6 & 11 &  0.019 &  63 &  7 & & & 6 & 10 & 12 &  0.016 &  36 &  3 \\
  3 &  7 & 10 &  0.018 &  89 & 11 & & & 8 &  9 & 12 &  0.015 &  36 &  3 \\
  7 & 10 & 11 &  0.018 &  69 & 12 & & & 9 & 10 & 12 &  0.017 &  42 &  2 \\

\hline
\end{tabular} \\

\begin{tabular}{c}
Quadratic Solutions
\end{tabular} \\
\begin{tabular}{cccccccccccccc}\hline
 & \multicolumn{4}{c}{B-spectra} & & & & & \multicolumn{4}{c}{B- and  V-spectra} &  \\
\hline
L1 & L2 & L3 & $r_{av}$ & N$_u$ & N$_e$ &  &  & L1 & L2 & L3 & $r_{av}$ & N$_u$ & N$_e$  \\
\hline

   1 &  4 &  6 &  0.018 &  56 &  2 & & &  1 &  8 & 11 &  0.012 &  27 &  1 \\
   1 &  8 & 11 &  0.012 &  27 &  1 & & &  1 &  8 & 12 &  0.015 &  27 &  1 \\
   3 &  4 & 10 &  0.017 &  91 &  9 & & &  3 &  7 & 10 &  0.017 &  89 & 11 \\
   3 &  5 & 11 &  0.018 &  63 &  7 & & &  3 & 10 & 12 &  0.016 &  37 &  2 \\
   3 &  7 & 10 &  0.017 &  89 & 11 & & &  4 & 10 & 12 &  0.014 &  38 &  1 \\
   3 &  8 & 10 &  0.018 &  92 &  8 & & &  5 &  8 & 12 &  0.016 &  38 &  1 \\
   3 &  9 & 10 &  0.017 &  92 &  8 & & &  6 &  8 & 12 &  0.015 &  38 &  1 \\
   5 &  8 & 11 &  0.018 &  71 & 10 & & &  6 & 10 & 12 &  0.017 &  38 &  1 \\
   6 &  9 & 10 &  0.019 & 105 &  6 & & &  7 &  8 & 12 &  0.014 &  38 &  1 \\
   9 & 10 & 11 &  0.017 &  82 &  4 & & &  8 & 10 & 12 &  0.013 &  38 &  1 \\

\hline
\multicolumn{12}{c}{\small\sl N$_u$:used stars, N$_e$:eliminated stars} \\ 

\end{tabular}
\end{center}
\end{table}

\begin{table}
\caption{Frecuency N with which lines L were used for the Best Mv-solutions}
\begin{center}
\begin{tabular}{cccccc}\hline\hline

\multicolumn{2}{c}{Mv-solutions} & & & \multicolumn{2}{c}{$(B-V)_0$-solutions} \\
\hline

L & N & & & L & N \\
\hline

 1 &  8 & & &  1 &  8 \\
 2 & 13 & & &  2 &  3 \\
 3 &  7 & & &  3 & 18 \\
 4 & 23 & & &  4 &  5 \\
 5 &  2 & & &  5 &  6 \\
 6 &  4 & & &  6 &  9 \\
 7 &  4 & & &  7 &  6 \\
 8 &  6 & & &  8 & 16 \\
 9 & 26 & & &  9 &  5 \\
10 &  6 & & & 10 & 20 \\
11 & 22 & & & 11 & 10 \\
12 & 14 & & & 12 & 17 \\
\hline

\end{tabular}
\end{center}
\end{table}

\newpage
\begin{figure}[!t]
\includegraphics[height=\columnwidth,angle=-90]{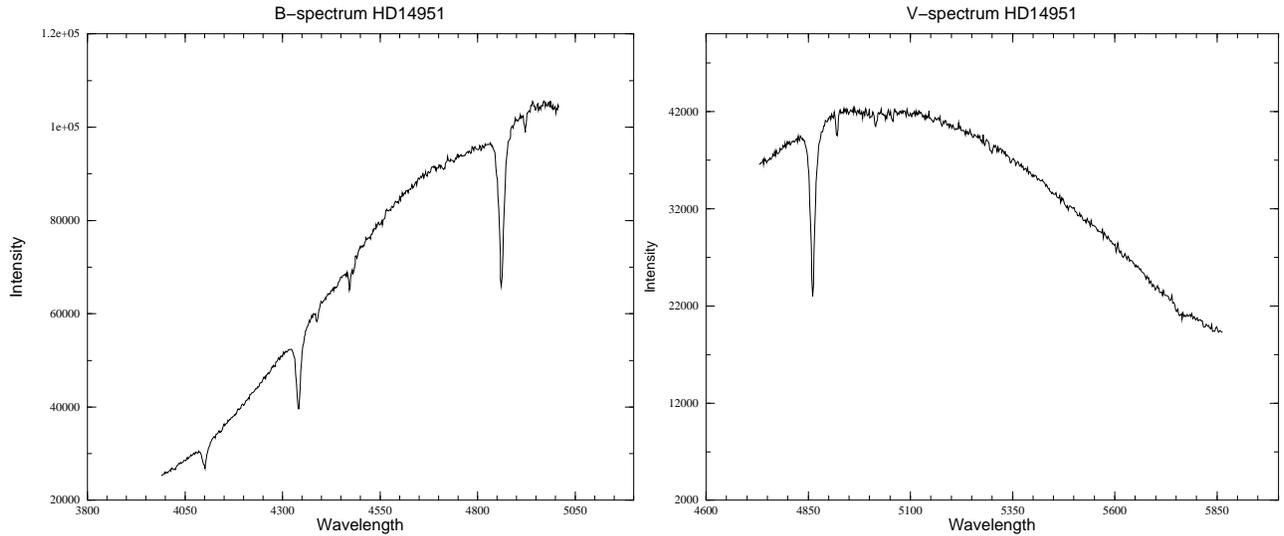}
\caption{B- and V-type observed spectra of HD14951}
\end{figure}

\begin{figure}
\includegraphics[height=\columnwidth,angle=-90]{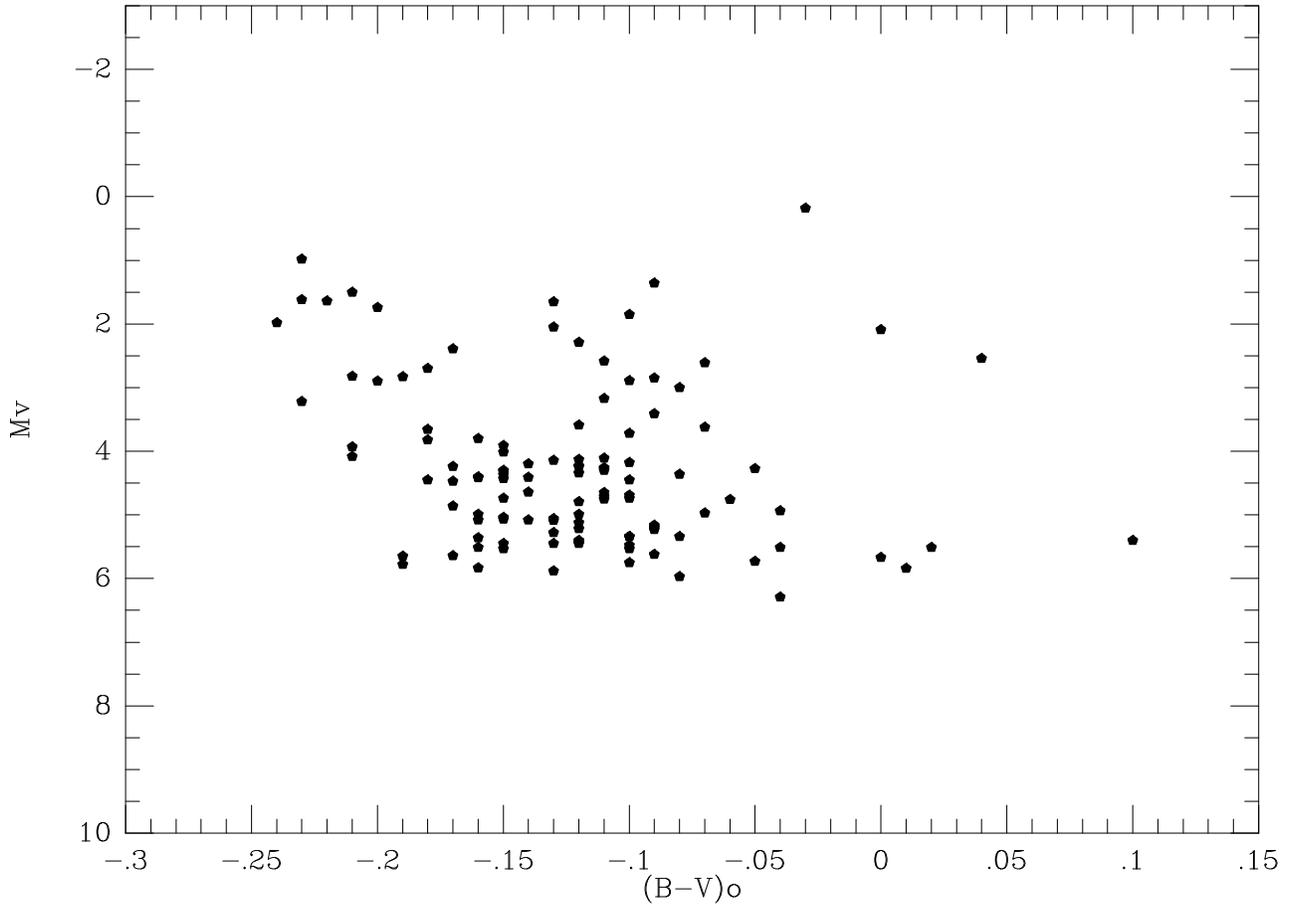}
\caption{H-R Diagram of observed stars}
\end{figure}

\newpage
\begin{figure}[!t]
\includegraphics[height=\columnwidth,angle=-90]{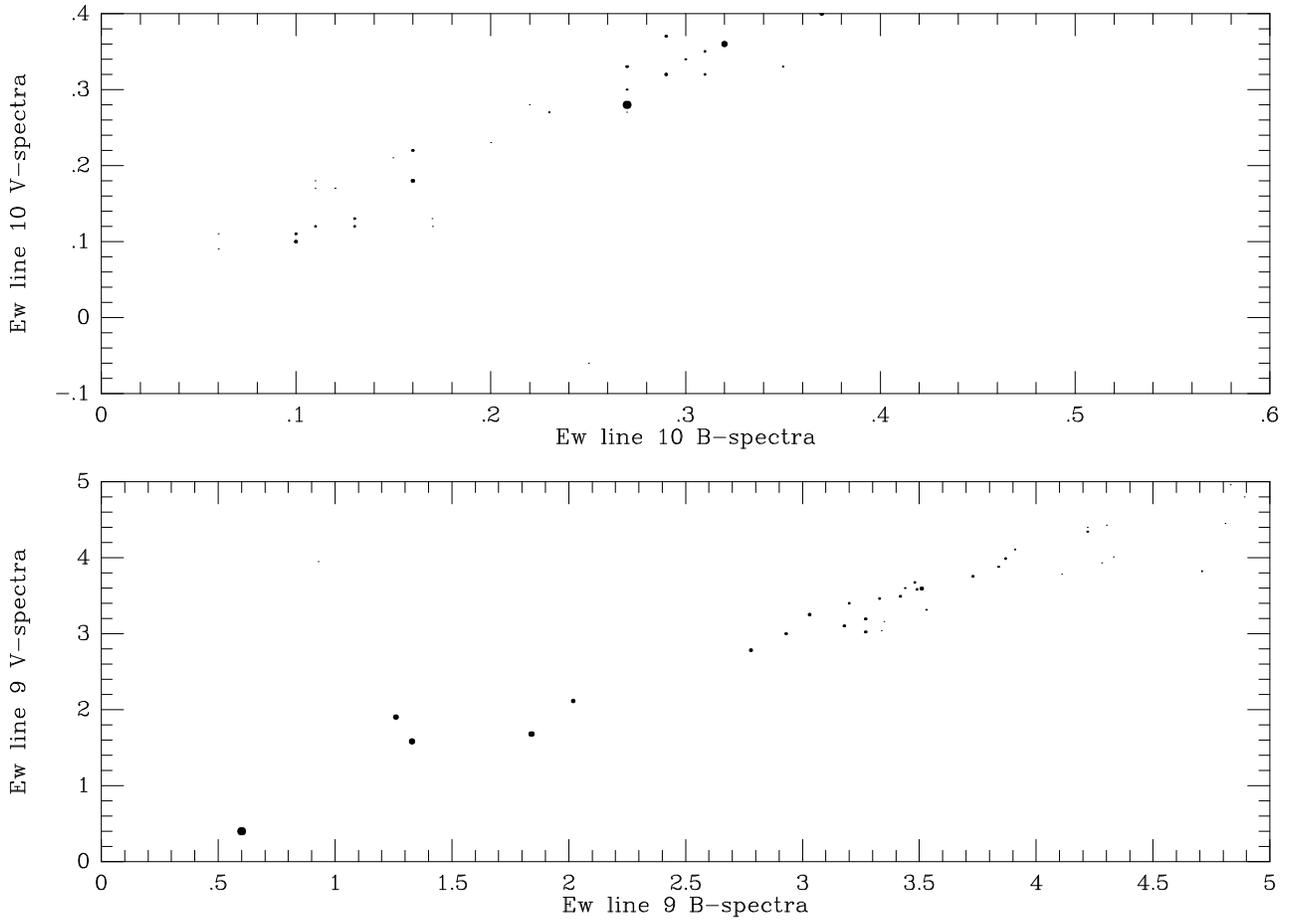}
\caption{Relation between the equivalent widths Ew measured in the B- and V-spectra}
\end{figure}

\begin{figure}
\includegraphics[height=\columnwidth,angle=-90]{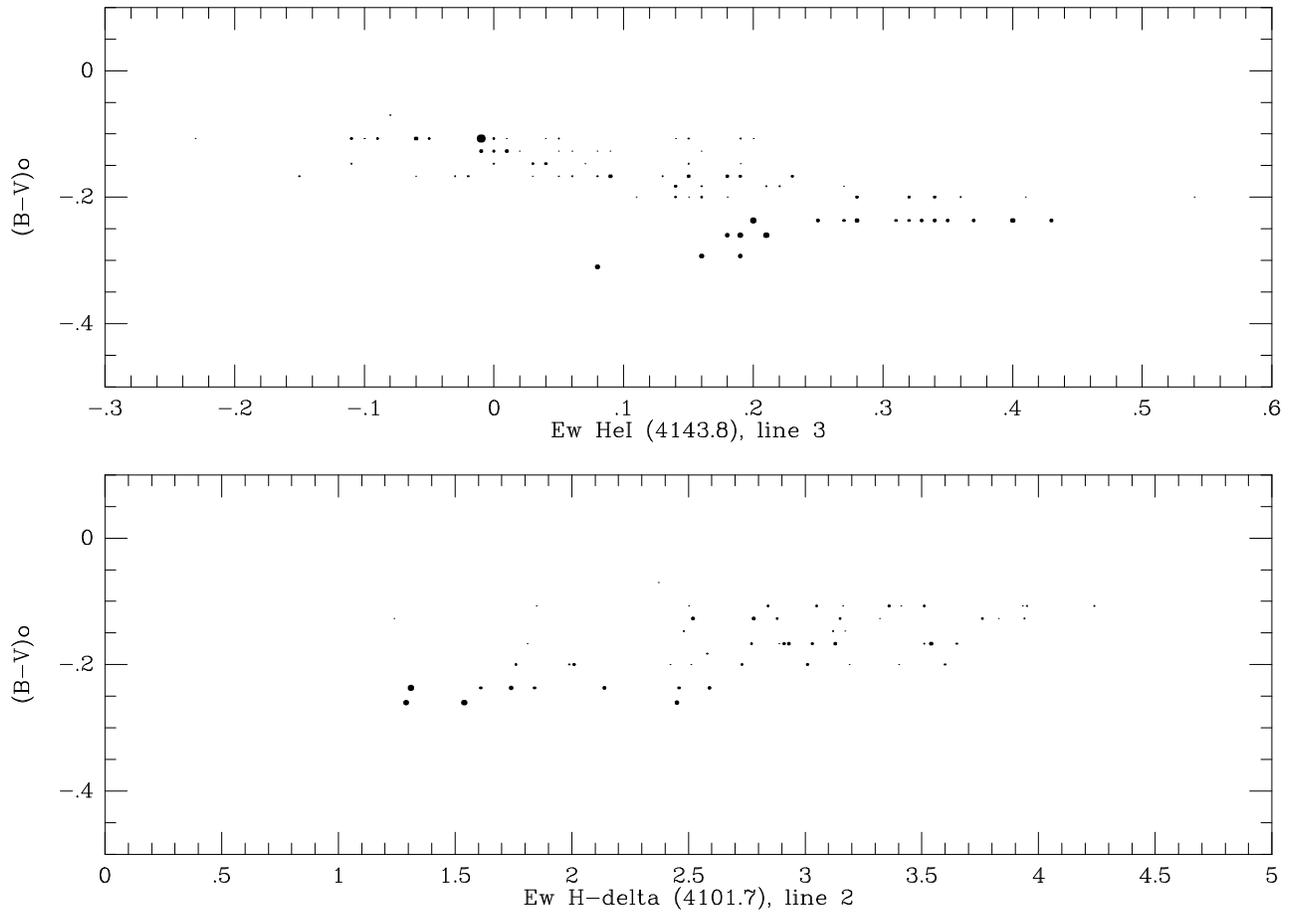}
\caption{Relation between the intrisic colors $(B-V)_0$ and the equivalents widths
	Ew for H-$\delta$ and HeI(4143)}
\end{figure}

\end{document}